\documentclass[12pt,twoside,fleqn]{article}
\usepackage{epsfig}
\usepackage{psfig}
\def\lsim{\raise0.3ex\hbox{$<$\kern-0.75em\raise-1.1ex\hbox{$\sim$}}}
\def\gsim{\raise0.3ex\hbox{$>$\kern-0.75em\raise-1.1ex\hbox{$\sim$}}}
\makeatletter
\@addtoreset{equation}{section}
\makeatother

\newcommand{\beqn} {\begin{equation}}
\newcommand{\eqn} {\end{equation}}

\newcommand{\slsh}[1] {#1\kern-.43em/}
\newcommand{\real}{{\sf I}\kern-.12em{\sf R}}
\newcommand{\comp}{{\sf I}\kern-.48em{\sf C}}
\newcommand{\nin} {\in\kern-.6em/}
\newcommand{\Tr} {\mbox{Tr}}

\def\MEF{m_{\rm eff}}\def\mef{\ifmmode\MEF\else$\MEF$\fi}

\def\AP{{Ann.\ Phys.\ }}

\def\LAT{{\NP B (Proc.\ Suppl.)\ }}
\def\NP{{Nucl.\ Phys.\ }}

\def\PL{{Phys.\ Lett.\ }}
\def\PR{{Phys.\ Rev.\ }}
\def\PRL{{Phys.\ Rev.\ Lett.\ }}

\begin{document}
\thispagestyle{empty}
%
%\hbox{}
 \mbox{} \hfill BI-TP 95/21\\
 \mbox{} \hfill FSU-SCRI-95-53\\
 \mbox{} \hfill May 1995\\
\begin{center}
\vspace*{1.0cm}
{{\large \bf The Gluon Propagator at High Temperature}
 } \\
\vspace*{1.0cm}
{\large U.M. Heller$^1$, F. Karsch$^2$, J. Rank$^2$} \\
\vspace*{1.0cm}
{\normalsize
$\mbox{}$ {$^1$ SCRI, Florida State University, Tallahassee, FL 32306-4052, USA\\
$^2$ Fakult\"at f\"ur Physik, Universit\"at Bielefeld, P.O. Box 100131,
D-33501 Bielefeld, Germany}}\\
\vspace*{2cm}
{\large \bf Abstract}
\end{center}
\setlength{\baselineskip}{1.3\baselineskip}

We study the gluon propagator in Landau gauge in the deconfined phase
of $SU(2)$ gauge theory. From the long-distance behaviour of correlation
functions of temporal and spatial components of
the gauge fields we extract electric ($m_e$) and magnetic ($m_m$)
screening masses. For temperatures larger than twice $T_c$ we find no
additional temperature dependence in $m_e(T)/T$, while $m_m(T)/T$
drops with increasing temperature. The decrease is consistent with
the expected behaviour, $m_m(T) \sim g^2(T)T$. We find $m_e(T) = 2.484(52)T$
and $m_m(T) = 0.466(15) g^2(T) T$.

\newpage
\setcounter{page}{1}
A basic non-perturbative feature of the high temperature plasma phase of QCD
is the occurrence of electric and magnetic screening masses for the gluon.
They play an important role in controlling the infrared behaviour of QCD
\cite{Lin80}.
While the electric screening mass can be calculated in leading order perturbation
theory and is found to be $O(gT)$, little is known about the magnitude of
the magnetic screening mass, which is expected to be $O(g^2T)$ \cite{Lin80}.
However, also other functional dependences have been obtained
in approximate non-perturbative approaches \cite{Kaj82,Kal92}. Recently
some attempts have been made to determine the magnetic screening mass
through the analysis of gap equations \cite{Esp93,Buc94,Phi94}.

It is the purpose of this paper to discuss a detailed study of the
temperature dependence of the magnetic screening mass, $m_m(T)$, obtained
from calculations of the finite temperature gluon propagator on the lattice.
Early lattice studies have attempted to extract this quantity from gauge
invariant observables, using lattices with twisted boundary conditions
\cite{Bil81,DeG82}.
Likewise the electric screening mass usually is extracted from gauge
invariant Polyakov loop correlation functions. Here we will determine the
effective gluon masses directly from the long-distance behaviour of the
gluon propagator. Although this requires the fixing of a gauge and leads to
a gauge dependent definition of a screening mass, it has the
advantage that it is closest to the perturbative definition of these masses
and allows a direct comparison with perturbative calculations. A more
detailed analysis of the momentum space representation of the gluon
propagator may, however, also lead to the determination of a gauge
independent pole mass \cite{Reb93}. 

We have performed calculations on large lattices of size $N_\sigma^3 \times
N_\tau$. In most cases we use $N_\tau = 8$, which insures that the
calculations in the high temperature phase are performed at gauge couplings
well inside the scaling region of the $SU(2)$ lattice gauge theory, i.e.
$\beta \ge 2.6$. For the spatial lattice we generally use $N_\sigma =32$,
which allows us to analyze correlation functions up to distances $zT=2$.
We have performed calculations in a large temperature interval
from $T\simeq 1.3T_c$ up to
$T \simeq 16T_c$ in order to become sensitive to possible logarithmic
corrections to the leading linear dependence of the high temperature
screening masses on $T$. Typically we have generated 800 independent gauge
field configurations on which we fix the Landau gauge,
$|\partial_{\mu}A^{\mu}({\bf x})|^2 = 0$, in order to analyze
gluon correlation functions. For the gauge fixing we
use a combination of overrelaxation and FFT algorithms \cite{Dav88,Man90}.
We have discussed the performance of this algorithm in Ref.~\cite{Kar95}.

We define the gauge fields, $A^{\mu}({\bf x})$, in the usual way 
from the $SU(2)$ link variables, $U_\mu({\bf x})$, the dynamical variables
in the lattice formulation,
\begin{equation}
A_{\mu}({\bf x}) = \frac{-i}{2g}
\left( U_{\mu}({\bf x}) - U^{\dagger}_{\mu}({\bf x}) \right) \ .
\label{a_mu}
\end{equation}
In a perturbative context the electric and magnetic screening masses are defined
through the zero momentum limit of the gluon polarization tensor,
$\Pi_{\mu \mu}(p_0,\vec{p})$, at vanishing Matsubara frequency $p_0$. In
coordinate space this limit can be realized through the long distance
behaviour of correlation functions of gauge field operators in one of the
spatial directions of the lattice, e.g. in the $z$-direction,
\begin{equation}
G_\mu(z) = \left\langle \Tr \; A_\mu (z)A_\mu (0) \right\rangle~~,
\end{equation}
where the fields $A_\mu (z) = \sum_{x_0,x_1,x_2}A_\mu(x_0,\vec{x})$ are
obtained by averaging over
a hyperplane transverse to the $z$-direction. In this way we project onto
$p_0=p_1=p_2=0$ and the limit $p_3\rightarrow 0$ corresponds to $z\rightarrow
\infty$. The long-distance behaviour of
$G_e(z) \equiv G_0(z) \sim \exp\{-\tilde{m}_e z\}$
is related to the electric screening mass, while
$G_m(z) \equiv 0.5(G_1 (z)+G_2 (z)) \sim \exp\{-\tilde{m}_m z\}$
yields the magnetic mass\footnote{We denote by $\tilde{m}$ masses in lattice
units. As all calculations, except one, have been performed on lattices with
$N_\tau =8$ sides
in the temporal direction, the masses in units of the temperature are given by
$m/T=N_\tau\tilde{m}=8\tilde{m}$.}. Due to the Landau gauge condition, it is 
easy to see that $G_3(z) =$~const. \cite{Man88}.

A problem which may arise with the fixing of the Landau gauge is the occurrence of
the so-called Gribov ambiguity \cite{gribov}. The Landau gauge condition 
$|\partial_{\mu}A^{\mu}({\bf x})|^2 = 0$ is realized on each lattice configuration
by maximizing
% the real part of 
the trace of the link fields, $U_\mu(x)$,  
\begin{equation}
\Sigma = \sum_{\mu, x} \Tr \, \left[ U_{\mu}(x) + U^{\dagger}_{\mu}(x)
\right]~~,
\label{spur_summe_1b}
\end{equation}
in the space of gauge equivalent fields \cite{Dav88,Man90,Kar95}. This
maximization procedure, however, is not unique. If one performs on an ungauged
configuration first a random gauge transformation and then maximizes
$\Sigma$, it is possible that one ends in a different 
Gribov sector. To investigate the dependence of $m_e(T)$ and
$m_m(T)$ on Gribov copies, we took 100 configurations, created 
on each of them 25 random gauge copies, and then performed the gauge fixing.  
Indeed, we find different
Gribov copies, which are distinguished by the value of $\Sigma$. In order to
test the sensitivity of $m_e(T)$ and $m_m(T)$ on this we selected the two samples
of 100 configurations with smallest and largest value of $\Sigma$,
respectively. On each sample we calculated the screening masses from the
exponential decay of the correlation functions $G_e$ and $G_m$. Within our
statistical accuracy we do not see any dependence of the screening masses on
the different Gribov sectors.   

\begin{figure}[htb]
\begin{center}
\vskip -1.9truecm
   \epsfig{bbllx=500,bblly=350,bburx=-35,bbury=60,
       file=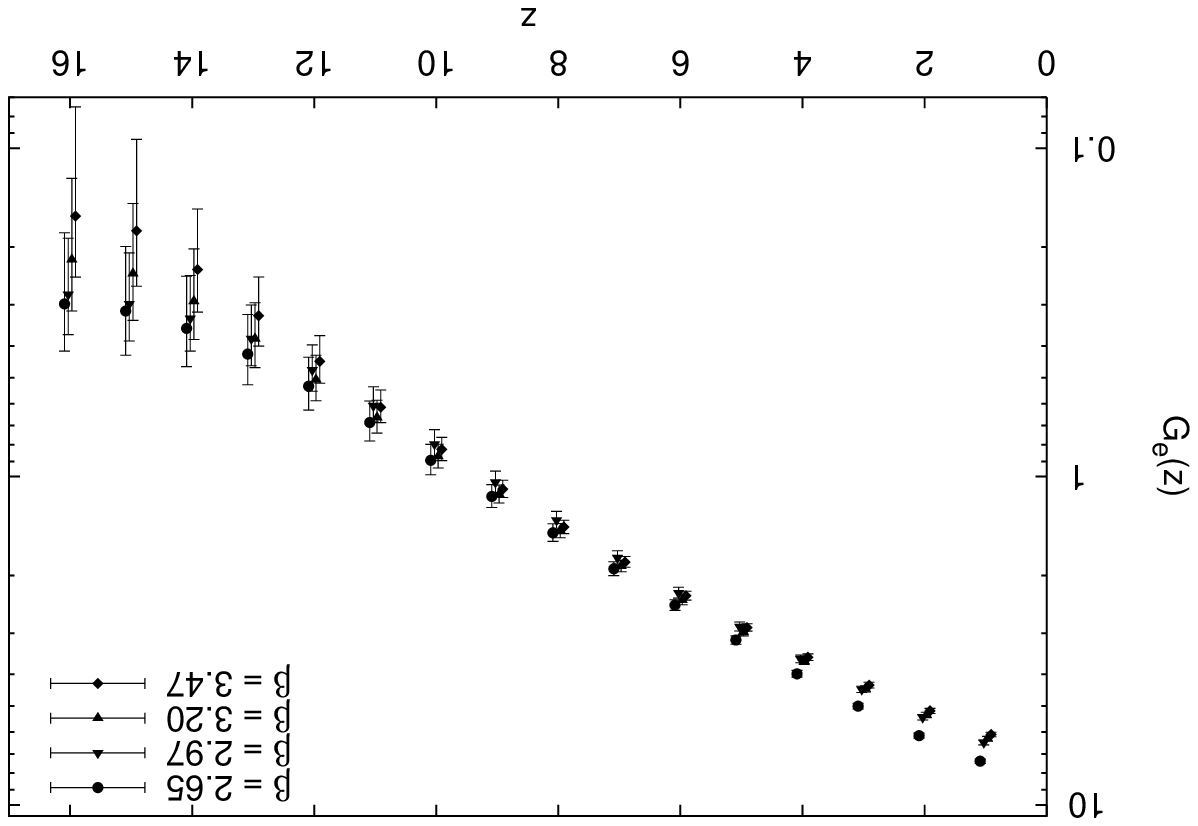,width=140mm,angle=-180}
\vskip -1.051truecm
   \epsfig{bbllx=500,bblly=350,bburx=-40,bbury=60,
       file=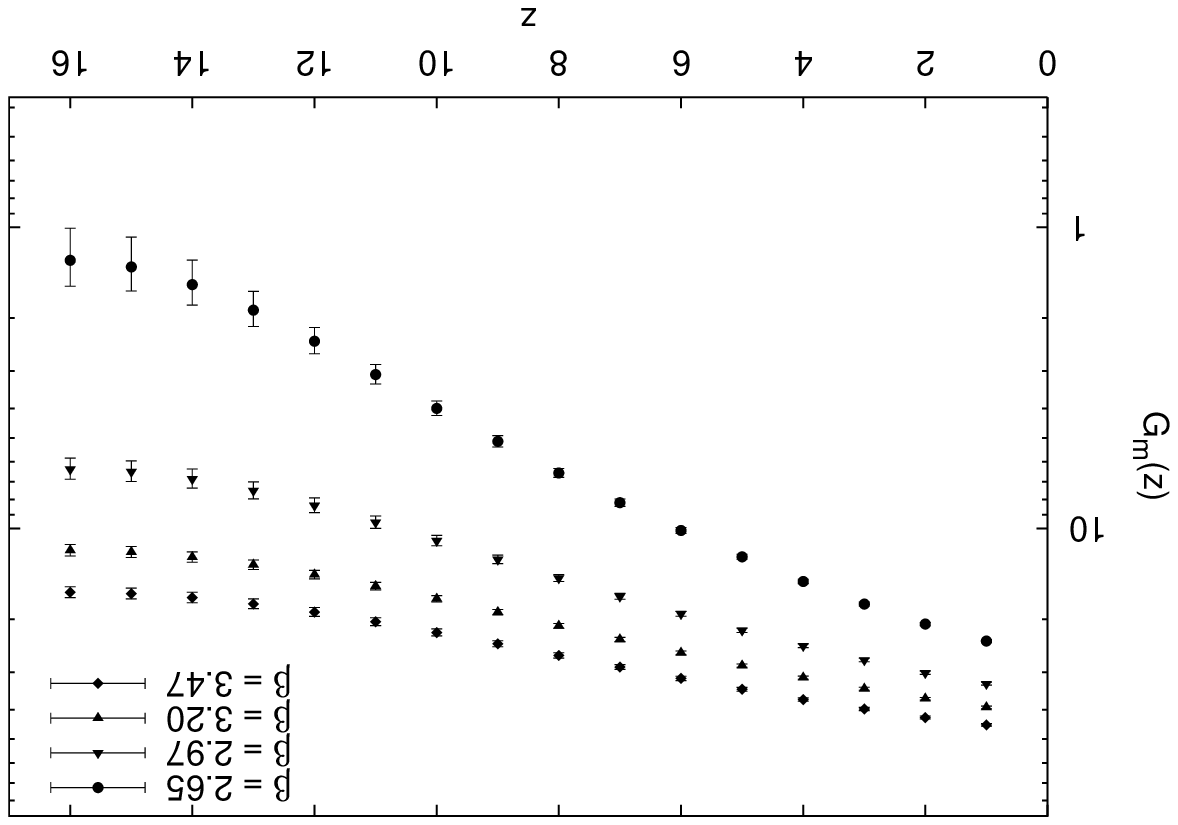,width=140mm,angle=-180}
\end{center}
\vskip -0.35truecm
\caption{The electric, $G_e(z)$, and magnetic, $G_m(z)$, correlation
functions for various values of the gauge coupling as a
function of $z$ calculated on lattices of size $32^3 \times 8$. The
temperatures corresponding to the different $\beta$-values can be found
in Table 1. They cover the interval $T/T_c \in [1.5,16]$.
The data points in the upper figure have been displaced horizontally
for better viewing.}
\label{fig:ln_prop}
\end{figure}
In Fig.~\ref{fig:ln_prop} we show the electric and magnetic gluon propagators at
various temperatures. Already from this figure it is obvious that the two
correlation
functions lead to quite different temperature dependences
of $m_m/T = 8\tilde{m}_m$ and $m_e/T = 8 \tilde{m}_e$. While $m_e/T$ does
seem to be temperature independent, the magnetic mass clearly rises slower
than linear in $T$, i.e. the correlation function $G_m(z)$ becomes
flatter with increasing temperature, which suggests a decrease of $m_m/T$.

\begin{figure}[htb]
\begin{center}
\vskip -1.8truecm
   \epsfig{bbllx=500,bblly=350,bburx=-35,bbury=60,
       file=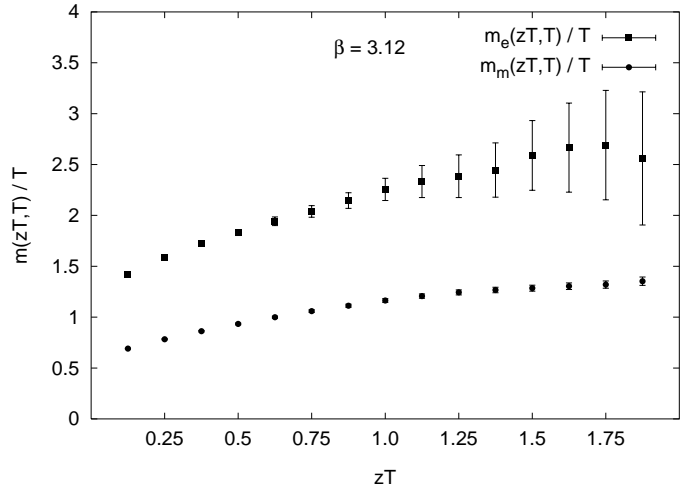,width=140mm,angle=-180}
\end{center}
\caption{The local electric and magnetic screening masses
for $\beta = 3.12$  in the range $zT \in [0.125,1.875]$, i.e. $1\le z \le 15$.}
\label{fig:lok_massen}
\end{figure}

We have analyzed the long-distance behaviour of the correlation functions by
studying the behaviour of local masses,
\begin{equation}
\frac{G_i(z)}{G_i(z+1)} =
\frac{{\rm cosh} \left( \tilde{m}_i \left( z - \frac{N_{\sigma}}{2}
\right) \right)}
{{\rm cosh} \left( \tilde{m}_i \left( z + 1 - \frac{N_{\sigma}}{2}
\right) \right)} \quad , \quad i=e,~m
\label{mratios}
\end{equation}
as well as by fitting the correlation functions with the ansatz, $G_i(z) =
c\; {\rm cosh} ( \tilde{m}_i ( z - N_{\sigma}/2))$. From an analysis of the local
masses we find that they slowly approach a plateau for $zT > 1$.
This behaviour is shown in Fig.~\ref{fig:lok_massen} for $\beta = 3.12$.

\begin{figure}[htb]
\begin{center}
%\leavevmode
\vskip -1.8truecm
   \epsfig{bbllx=500,bblly=350,bburx=-35,bbury=60,
       file=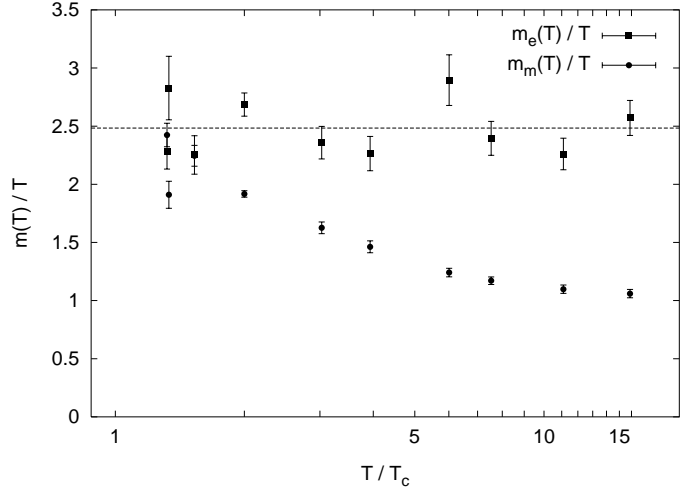,width=140mm,angle=-180}
\end{center}
%\vskip -0.5truecm
\caption{Electric and magnetic screening masses in units of the
temperature versus $T/T_c$. The screening masses have been extracted
from fits to the gluon correlation functions for distances $zT \ge 1$.
The dashed line shows the fit to the electric mass given in Eq.(\ref{mefit})}
\label{fig:massen_ab_8}
\end{figure}

Also the single-cosh fits yield acceptable $\chi^2$-values only, if the short
distance part of the correlation functions ($zT < 1$) is left out from the
fit. This is obvious from Fig.~\ref{fig:ln_prop}. We generally observe that
a fit to the propagator for $zT>1$ yields masses which are consistent
within statistical errors with the local masses extracted at $zT=1$ ($z=8$).
In Table~\ref{massen_daten_ab_8} we give the results from our fits at all values
of the gauge coupling analyzed. These results are also shown in
Fig.~\ref{fig:massen_ab_8}. In order to relate the gauge coupling used in the
calculation to a temperature we make use of a parameterization of asymptotic
scaling violations of the $SU(2)~~\beta$-function \cite{Eng95}.

\begin{table}
\begin{displaymath}
\begin{array}[t]{| l | r || l | l |} \hline
% \rule[-0.6em]{0em}{1.7em} \multicolumn{1}{c |}{\beta} &
\multicolumn{1}{ |c |}{\beta} & \rule[-0.6em]{0em}{1.7em} T/T_c &
\multicolumn{1}{ |c |}{\frac{m_e(T)}{T}} &
\multicolumn{1}{ |c |}{\frac{m_m(T)}{T}} \\ \hline\hline
2.60       &  1.32 & 2.28(15) & 2.42(10) \\ \hline
2.74^{(1)} &  1.33 & 2.83(27) & 1.91(12) \\ \hline
2.65       &  1.53 & 2.25(17) & 2.25(9) \\ \hline
2.74^{(2)} &  2.00 & 2.69(10) & 1.92(3) \\ \hline
2.88       &  3.03 & 2.36(14) & 1.63(5) \\ \hline
2.97       &  3.93 & 2.26(15) & 1.46(5) \\ \hline
3.12       &  6.01 & 2.90(22) & 1.24(4) \\ \hline
3.20       &  7.53 & 2.40(15) & 1.17(3) \\ \hline
3.34       & 11.10 & 2.26(14) & 1.10(4) \\ \hline
3.47       & 15.88 & 2.57(15) & 1.06(4) \\ \hline
\end{array}
\end{displaymath}
%\vspace{-3ex}
\caption{Electric and magnetic screening masses in units of the temperature,
$m_e(T)/T$ and $m_m(T)/T$ for temperatures
$1.32 \le T/T_c \le 15.88$. Calculations have been performed on lattices of
size $32^3 \times 8$ except for the cases (1) and (2), where calculations
have been performed on a $ 32^3 \times 12$~(1) and
a $32^2 \times 64 \times 8$~(2) lattice. The masses have been obtained from
fits to the correlation functions for distances $zT\ge 1$.}
\label{massen_daten_ab_8}
\end{table}
We clearly see that the electric screening mass is proportional to the
temperature. Up to temperatures of $16T_c$ we do not see any indication
for the expected perturbative behaviour $m_e \sim g(T) T$. Averaging over
the results for $T \ge 2T_c$ we obtain
\begin{equation}
m_e (T) = (2.484 \pm 0.052) T~~.
\label{mefit}
\end{equation}
This agrees with the findings of a recent analysis of Polyakov loop
correlation functions, where the electric mass has been determined with
the help of the
transfer matrix approach \cite{engels5}. We will return to
a more detailed discussion of the electric mass after having analyzed
the magnetic screening mass.

The magnetic screening mass clearly rises slower than
proportional to $T$. We thus may compare the results with the expected
behaviour $m_m \sim g^2(T) T$. In Fig.~\ref{fig:g_2_ab_8} we show the
inverse mass, $T/m_m(T)$. This should rise logarithmically, if the
gauge coupling is running according to the leading orders of the
renormalization group equation, $g^{-2}(T) \sim \ln(T/T_c)$. We note that
this does indeed describe our numerical results quite well.
The spatial string tension is also expected to be
sensitive to magnetic screening in the high temperature phase and indeed
a similar scaling behaviour has been observed for it in \cite{Bal93}.

\begin{figure}[htb]
\begin{center}
%\leavevmode
\vskip -1.8truecm
   \epsfig{bbllx=500,bblly=350,bburx=-35,bbury=60,
       file=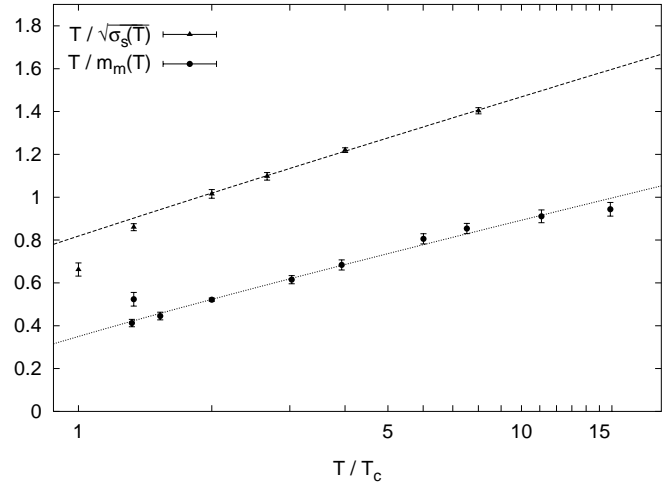,width=140mm,angle=-180}
\end{center}
%\vskip -0.5truecm
\caption{Inverse magnetic mass and square root of the spatial string
tension (from Ref.~[17]) versus $T/T_c$. The lines show the fits given in
Eqs.(\ref{mmfit}) and (\ref{sigma_s}).}
\label{fig:g^2_ab_8}
\end{figure}
We have fitted the magnetic mass with the ansatz
\begin{equation}
{T \over m_m(T)} = c g^{-2} (T)~~,
\label{mmfit_ansatz}
\end{equation}
assuming the validity of the two-loop formula for $g^2(T)$ with a free
$\Lambda$ parameter,
\begin{equation}
g^{-2}(T) = {11 \over 12 \pi^2} \ln{T/\Lambda_m} +
          {17 \over 44 \pi^2}  \ln(2\ln{T/\Lambda_m})~~~~.
\label{twoloop}
\end{equation}
{}From a fit of the numerical results for $T \ge 2T_c$ we find
\begin{equation}
m_m (T) = (0.466 \pm 0.015) g^2(T) T ~~~,
\label{mmfit}
\end{equation}
with $\Lambda_m = 0.262(18) T_c$.
We note that our result for the magnetic mass is about twice as large
as the earlier numerical results \cite{Bil81,DeG82}. This may not be too
surprising in view of the rather small lattices used in these first
studies\footnote{The analysis performed in Ref.~\cite{Bil81} could only probe
distances $zT\le 0.5$ and the analysis in Ref.~\cite{DeG82} has been
performed at a value of the gauge coupling which was too large to keep the
system confined in spatial directions.}. Our result also is
larger than analytic results based on a semiclassical approximation
\cite{Bir93} and gap equations \cite{Buc94,Phi94}.

It is interesting to relate the magnetic mass to the spatial string tension,
which also has been found to scale like $g^2T$ \cite{Bal93}. The
temperature dependent running coupling has been found to be somewhat smaller
in the case of $\sqrt{\sigma_s}$, which results from the different constant
terms that went into the definition of the $\Lambda$-parameter. We may,
however, express the result for the spatial string tension in terms of
the coupling
$g^2(T)$ determined here from the scaling of the magnetic mass. This yields
\begin{equation}
\sqrt{\sigma_s} = (0.368\pm 0.017) ~ \{1-(0.113\pm0.022)g^2(T)\} ~ g^2(T)T~~.
\label{sigma_s}
\end{equation}
Combining this with Eq.~(\ref{mmfit}) we find
\begin{equation}
m_m (T) = (1.27\pm 0.10) ~ \{1+(0.113\pm0.022)g^2(T)\} ~ \sqrt{\sigma_s}~~.
\label{compare}
\end{equation}
The spatial string tension has also been found to be quantitatively
closely related to the string tension, $\sqrt{\sigma_3}$, of
the three-dimensional $SU(2)$
gauge theory. We thus can use the above relation to compare the result for
the magnetic mass of the (3+1)-dimensional $SU(2)$ gauge theory at finite
temperature with the mass gap of the three-dimensional $SU(2)$ theory.
The latter is found to be about $5\sqrt{\sigma_3}$ \cite{Tep92},
i.e. (4-5)-times larger than the magnetic mass found here. It will be
interesting to also analyze, in the future, the effective gluon mass in the
three-dimensional gauge theory within the approach discussed here. This will
clarify whether the gluon masses can be interpreted as constituent masses of
the glueballs and to what extend the thermal gluon mass in (3+1) dimensions and 
the gluon mass in three dimensions are related in a similar way as
$\sigma_s(T)$ and $\sigma_3$.

In view of the discussion of the magnetic mass we shall now return to the
result for the electric mass given in Eq.~(\ref{mefit}). The running coupling
extracted from the magnetic mass varies by a factor two in the temperature
interval studied by us, $g^2(2T_c)\simeq 4.1$ and $g^2(16T_c)\simeq 2.2$
(similar values have been obtained from $\sqrt{\sigma_s}$). From the leading
order perturbative result, $m_e(T) = \sqrt{2/3} g(T) T$, we thus would expect a
30\% drop of $m_e/T$ in this temperature interval. Moreover, using the
perturbative form at our largest temperature, we would conclude that the
coefficient in front of $g(T)T$ comes out to be a factor two larger than
suggested by leading order perturbation theory. Such an enhancement is also
found when higher order effects, including the contribution from a
non-vanishing mass, are taken into account through resumed perturbation
theory \cite{Reb93}. It thus seems that indeed non-perturbative effects give
large contributions to the electric screening mass and eliminate the leading
$g(T)$-dependence. It will be interesting to perform calculations at even
higher temperatures in order to see whether contact with leading order
perturbative behaviour can be made at all in the electric sector. Work in
this direction is in progress. 

\medskip
\noindent
{\bf Acknowledgements:}
The work of UMH was supported in part by the DOE under grants
\#~DE-FG05-85ER250000 and \#~DE-FG05-92ER40742 and the work of FK
in part by the NATO research grant CRG940451. 
The computations have been performed on Connection Machines at the HLRZ,
and SCRI. We thank the staff of these institutes for their support.

\end{document}